# Proposal of Michelson-Morley experiment via single photon interferometer: Interpretation of Michelson-Morley experimental results using de Broglie-Bohm picture

Masanori Sato

*Honda Electronics Co., Ltd.,*
*20 Oyamazuka, Oiwa-cho, Toyohashi, Aichi 441-3193, Japan*

E-mail: msato@honda-el.co.jp

**Abstract:** The Michelson-Morley experiment is considered via a single photon interferometer and we propose the interpretation of the Michelson-Morley experimental results using de Broglie-Bohm picture. We point out that the Michelson-Morley experiment revealed the interference of photons, however, it did not reveal the photons' simultaneous arrival at the beam splitter. According to the de Broglie-Bohm picture, the quantum potential nonlocally determines the interference of photons. The interference of the photons is not governed by the arrival time of the photons; rather, it is governed by the quantum potential. Therefore, there is a possibility for another interpretation of the Michelson-Morley experimental results.

1. INTRODUCTION

Although quantum mechanics and special relativity have a very interesting relationship, the concepts of these two theories appear to be quite different. We consider that quantum mechanics and special relativity can be discussed from the phenomenon of interference. We consider that quantum mechanics and special relativity both have, as a basis for their theories, experimental data on interference. This is the reason why they have a particular compatibility. The relationship between energy $\varepsilon$ and momentum $\mu$ as $\mu = \varepsilon/c$ (c: light speed) is one of the suitable examples.

We have previously pointed out that another means of interpretation is possible for Young's double slit experiment [1], and discussed quantum entanglement [2] and delayed choice experiment [3]. We have also discussed time delay in atomic clock in motion [4]. It is important to investigate these classical themes with respect to new understandings.

The Michelson-Morley experiment demonstrates that an interference pattern does not vary according to the motion of the earth. This is the fundamental data of the theory of special relativity. Thus the phenomenon of interference is critically related to the concept of special relativity, and is also related to quantum mechanics. For example, Young's double slit experiments demonstrate that interference is related to quantum mechanics. We consider that the phenomenon of interference is the key to establishing a conceptual relationship between quantum mechanics and special relativity.

Einstein's concept of special relativity, that is, all reference frames are equivalent, appears to be independent of the Michelson-Morley experiment. However, the theory of special relativity has a very good agreement with experimental data obtained through the Michelson-Morley experiment, and therefore, special relativity has a relationship with interference.

In this letter, we note that the Michelson-Morley experiment shows the interference of photons; however, it does not show the photons' simultaneous arrival at the beam splitter. The experiment also revealed that the interference conditions were not affected by the motion of the earth. The photon arrival should be investigated using photon detection technology. At this stage, we cannot distinguish the arrival time at the order of photon wavelength. However, this is simply due to technological limitation.

2. PROPOSAL

A. Single photon Michelson interferometer

**Figure 1** shows the schematic diagram of a single photon Michelson interferometer. A photon enters the interferometer via the beam splitter, is reflected by the mirror, and is then recombined by the beam splitter. We can detect the interference, that is, the photon paths can be arranged such that the detector detects the photons.

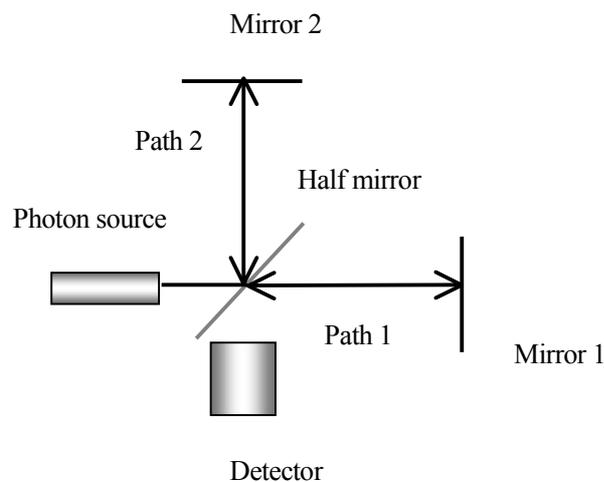

Fig. 1   Conceptual diagram of Michelson-Morley experiment

**Figure 2** shows a schematic diagram of the quantum potential. According to this schematic diagram, a single photon Michelson interferometer appears to detect only the interference, and does not measure the speed of photons. In a single photon interferometer, there is only a photon in the photon paths, therefore it cannot measure the arrival time of photons on path 1 and path 2. Only the interference condition is evaluated.

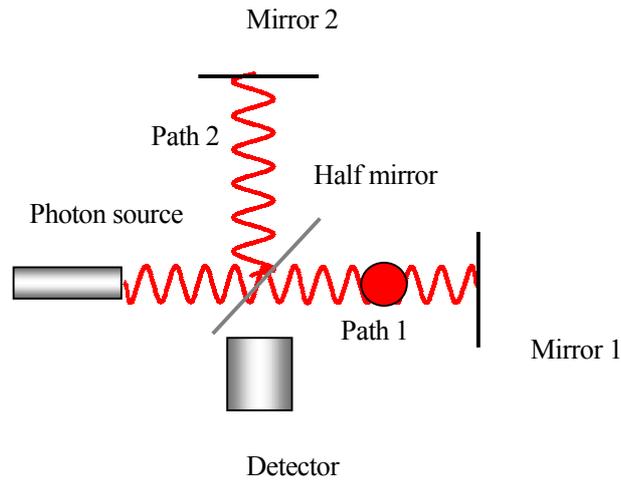

Fig. 2    Interpretation using de Broglie-Bohm picture

B. Sing around measurement of photon speed

In acoustic waves, for the measurement of sound velocity, not only interference but also the sing around method is used. The sing around method uses two pairs of transmitters and receivers as shown in **Fig. 3**, where a pulsed signal is transmitted by transmitter 1 and detected by detector 2, at the detection new pulsed signal is transmitted by transmitter 2 and detected by detector 1. We can detect the frequency of repetition which indicates the sound velocity. Using photons we can construct a setup for the sing around measurement of the photon velocity. Thus we can determine the photon velocity without the interference.

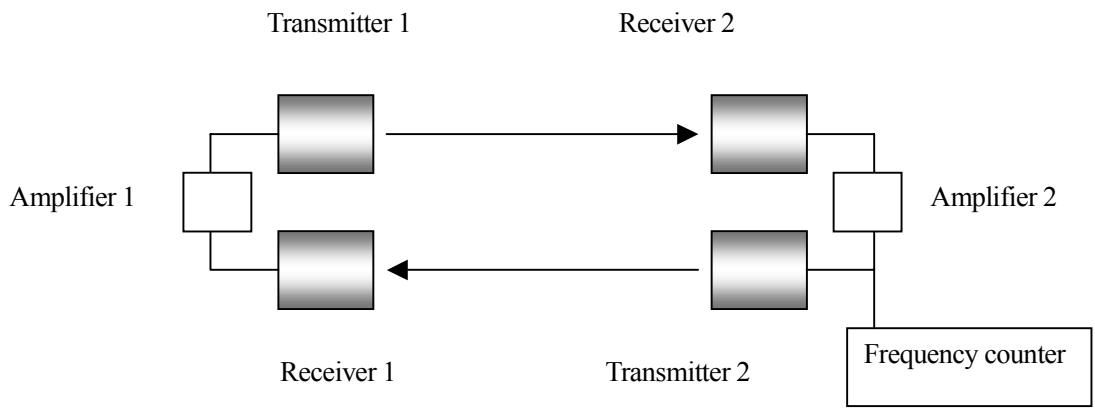

Fig. 3    Sing around sound source setup

**Figure 4** shows the Michelson-Morley type sing around measurement setup for determining the photon velocity. We can determine the photon velocity on the basis of the detection of the sing around frequency by using frequency counters 1 and 2 as the frequency changes. Thus the Photon speed can be measured directly without the interference.

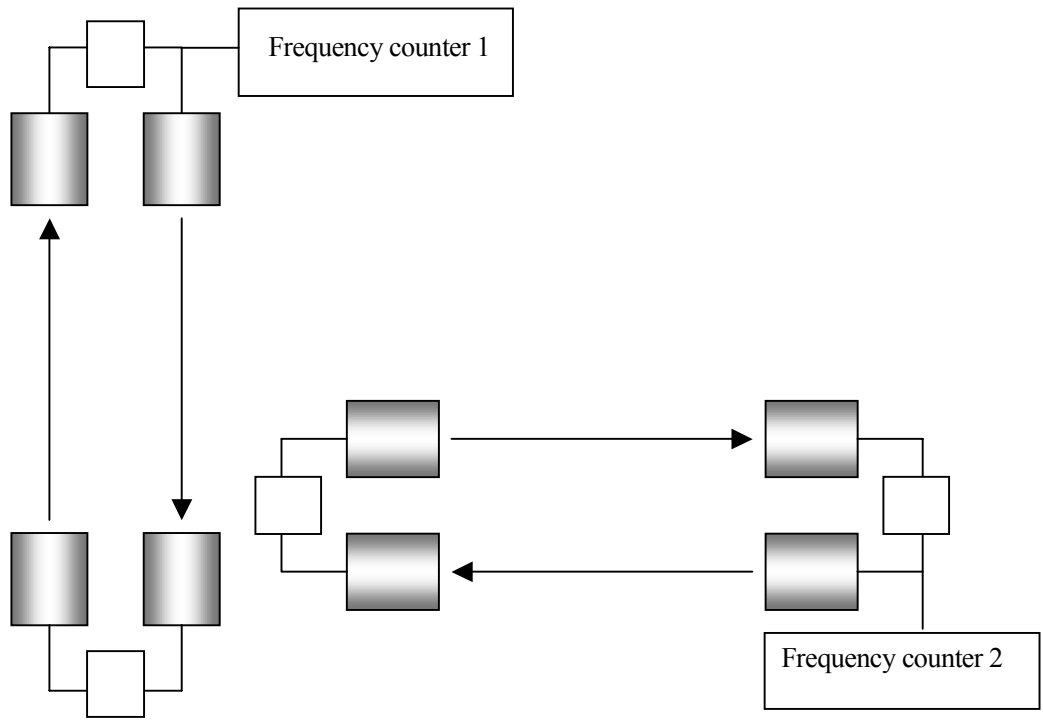

Fig. 4    Michelson-Morley type setup for the sing around measurement of photon velocity

3. DISCUSSION
A. Causality and special relativity

We define causality as the principle that in time, the cause must precede the effect, regardless distance. This is an old fashioned definition of causality prior to Einstein's proposition of special relativity. It means that we are operating in the absolute space frame instead of Einstein's time-space frame. We do not consider that causality requires a light cone. Of course, we accept that a photon transfers information by itself, namely, in terms of its flight of the photon itself, and therefore light speed is important for describing the relativity of positions A and B. Thus a light cone is functional with respect to the transfer of information by photons in this sense. However, during the interference, the photon in itself does not transfer any information on the interference.

Special relativity is defined as the principle of that all reference frames are equivalent. At this stage, we note that a phenomenon satisfied special relativity occurs through photons; this means that information is transferred by the photon itself, and that information between any two reference frames is restricted by photon speed. According to this definition, special relativity differs from causality. This implies that nonlocality does not have compatibility with special relativity, however does not break causality. This is because we define causality as the cause must precede the effect. Thus, nonlocality has compatibility with causality.

We have previously discussed causality, nonlocality and special relativity in quantum entanglement [2] and the interference in delayed choice experiment [3]. This is because quantum entanglement and interference do not appear to be the phenomena through photons. It is difficult to carry out the discussion of the interference of Michelson interferometer under special relativity using the finite light speed c, therefore, we try the de Broglie-Bohm picture.

B. Interpretation of Michelson-Morley experiment using de Broglie-Bohm picture

The de Broglie-Bohm picture [5, 6] provides a suitable means of interpreting this situation. The phenomenon is explained using the wave and particle model. The wave is the pilot wave (quantum potential) and the particle is a photon that is guided by the pilot wave [5, 6]. The wave is nonlocal and the particle, that is, the photon, is local. The photon travels at light speed, namely, the photon has compatibility with special relativity; however, the wave does not have compatibility with special relativity (i.e., nonlocal). Interference is determined by the geometry of the interferometer, as shown in **Fig. 2**, i.e., the interference is governed by the lengths of paths 1 and path 2. We point out that the Michelson-Morley experiment does not measure the velocity of the photon.

4. CONCLUSION

We discussed the Michelson-Morley experiment that employs a single photon interferometer, and pointed out that this experiment demonstrates that interference is independent of the motion of the earth. We also proposed a means of interpretation of this experiment using de Broglie-Bohm picture. We also pointed out that the direct measurement of photon velocity should be performed. In the future, we intend to carry out a modified Michelson-Morley experiment with the utilization of photon counting technology.